\documentclass[a4paper,12pt]{article}
\usepackage{epsfig}
\usepackage{amssymb}
\begin{document}
\thispagestyle{empty}

\newcommand{\etal}  {{\it{et al.}}}  
\def\Journal#1#2#3#4{{#1} {\bf #2}, #3 (#4)}
\def\PRD{Phys.\ Rev.\ D}
\def\NIMA{Nucl.\ Instrum.\ Methods A}
\def\PRL{Phys.\ Rev.\ Lett.\ }
\def\PLB{Phys.\ Lett.\ B}
\def\EPJ{Eur.\ Phys.\ J}
\def\IEEETNS{IEEE Trans.\ Nucl.\ Sci.\ }
\def\CPCD{Comput.\ Phys.\ Commun.\ }


\bigskip

{\Large\bf
\begin{center}
Critical point and conformal anomaly
\end{center}
}
\vspace{0.1 cm}

\begin{center}
{ G.A. Kozlov  }
\end{center}
\begin{center}
\noindent
 { Bogolyubov Laboratory of Theoretical Physics\\
 Joint Institute for Nuclear Research,\\
 Joliot Curie st., 6, Dubna, Moscow region, 141980 Russia  }
\end{center}
\vspace{0.1 cm}

 \begin{abstract}
 \noindent
 {We study the critical point (CP) phenomenon through the increase of fluctuations related to characteristic critical mode. CP would also be identified through the production of  primary photons induced by  conformal anomaly of strong and electromagnetic interactions. The novel approach to an approximate scale symmetry breaking is  developed for this.}


\end {abstract}




\bigskip

1. The heavy ion collisions at their very early stage produce the gluon fields that are abundant in the hot phase space at high temperatures $T$ and the baryon density. This is an ideal example to the theory with the conformal sector where the scale invariance is exact and no massive particles are emerging.  The invariance with respect to the scale transformations of coordinates $x_{\mu}\rightarrow \omega x_{\mu}$ ($\omega$ is an arbitrary constant) corresponds to the conservation of the dilatation current $S^{\mu} = \theta ^{\mu\nu} x_{\nu}$: $\partial_{\mu} S^{\mu} = \theta ^{\mu}_{\mu} = 0$, where $\theta^{\mu}_{\mu}$ is the trace of the energy-momentum tensor $\theta_{\mu\nu}$. 
Because of the presence of strong gluon fields the QCD vacuum is disordered and scale invariance is destroyed by the appearance of  the dimensional scale $\mu$  [1]
$$\mu = M_{UV}\,\exp (-8\,\pi^{2}/b_{0}\,g_{0}^{2}), $$
with $M_{UV}$ being the ultra-violet (UV) scale, $g_{0}$ is the bare coupling constant and $b_{0}$ is the coefficient in the $\beta$-function to be defined later. 
In the vicinity of QCD critical point (CP) 
$\mu\rightarrow 0$ at small enough $b_{0}$ and $g_{0}^{2}$.

Within the breaking of conformal invariance all the processes are governed by the conformal anomaly (CA) resulting from running of $g(\mu)$ in the $\beta$ -function. From this $S_{\mu}$ is already non-conserved in the theory containing, e.g., the gluon and quark degrees of freedom (d.o.f.): the divergence of $S_{\mu}$ is proportional to the $\beta$-function and the quark masses.  For $SU(N)$ gauge theory with $N$ number of colors and $N_{f}$ number of flavors in the fundamental representation, the $\beta$-function is
$$\beta(\alpha_{s}) \equiv \mu\frac{\partial\alpha_{s}(\mu)}{\partial\mu} \equiv -\frac{b_{0}}{2\pi}\alpha^{2}_{s} - \frac{b_{1}}{(2\pi)^{2}}\alpha^{3}_{s} + ...,$$
where $\alpha_{s} \equiv \alpha_{s} (\mu)$ is the renormalized gauge coupling constant defined at the scale $\mu$; $b_{0}$ and $b_{1}$ are known coefficients. There could be an approximate scale (dilatation) symmetry if $\beta(\alpha_{s})$ is small enough and $\alpha_{s}(\mu)$ is slowly running with $\mu$. Theory becomes conformal in the infra-red (IR)  with the non-trivial solution $\alpha_{s}^{\star} = -2\pi\,b_{0}/b_{1}$ (IR fixed point (IRFP) or the Banks-Zaks [2] conformal point) in the perturbative domain if $b_{0} = (11 N- 2 N_{f})/3$ is small. The latter is happened when $(N/N_{f}) \sim 2/11$. Once $N_{f}$ decreases near the phase transition (the value of $\alpha_{s}^{\star}$ increases) the CP is characterized by $\alpha^{c}_{s} < \alpha^{\star}_{s}$ at which the spontaneous breaking of the chiral symmetry is occurred, and the confinement does appear. In the neighborhood of the IRFP the $\beta$-function is approximated by [3]
$$\beta(\alpha_{s}) \simeq -\Delta\alpha_{s} (\mu/\Lambda)^{\delta}, $$
where $0 < \Delta\alpha_{s} = \delta\cdot (\alpha_{s}^{\star} - \alpha_{s}^{c}) << \alpha_{s}^{c}$, $\delta \leq O(1)$. The spontaneous breaking of chiral symmetry in certain gauge theories may also imply the spontaneous breaking of an approximate conformal symmetry [4]. Once the latter is broken spontaneously, a light scalar dilaton could emerge as a pseudo-Goldstone boson of scale symmetry.  

The very natural appearance of a dilaton comes from the field theory given by the Lagrangian density (LD) in the general form 
\begin{equation}
\label{e1}
L = \sum_{i} g_{i}(\mu)\,O_{i}(x),
\end{equation}
where the local operator $O_{i}(x)$ has the scaling dimension $d_{i}$. LD (\ref{e1}) will be scale invariant under dilatations $x^{\mu}\rightarrow e^{\omega} x^{\mu}$ with $O_{i} (x)\rightarrow e^{\omega d_{i}} O_{i}(e^{\omega} x)$ and $\mu\rightarrow e^{-\omega}\mu$ if $g_{i}(\mu)$ is replaced by $g_{i}(\mu)\rightarrow g_{i} [\mu (\chi/f)] (\chi/f)^{4-d_{i}}$. The new dilaton field $\chi(x)$ as the conformal compensator introducing a flat direction transforms  according to $\chi(x)\rightarrow e^{\omega}\chi (e^{\omega} x)$.  The order parameter $f= \langle\chi\rangle$ for scale symmetry breaking is dictated by the dynamics of a relevant strong sector. From this one obtains [5]
$$\theta^{\mu}_{\mu} = \sum_{i} g_{i}(\mu)\,(d_{i} - 4)\,O_{i}(x) + \sum_{i} \beta_{i} (g)\,\frac{\partial}{\partial g_{i}} L .$$ 
The theory is scale invariant if $d_{i} =4$ and $\beta_{i} (g) = 0$.
We suppose that at the scale $\geq\Lambda$ (QCD scale) the dilaton is formed as the bound state of two gluons, the glueball $\chi = O^{++}$, with the mass $m \sim O(\Lambda)$. The $\chi$ may be understood as the string ring solution  so that the ends of the string meet each other to form a circle with some finite radius.

2. The critical phenomena of strong interacting matter have been studied in many papers (see, e.g., the ref. in [6]). The mechanism towards the observation of CP via the influence quantum fluctuations of two-body Bose-Einstein correlations for observed particles to which the critical end mode couples has been  developed in [7].
In this paper, the CP is identified through the increase  of fluctuations related to some characteristic critical mode. The latter is naturally given by the  mass $ m$ of the dilaton with the correlation length $\xi = m^{-1}$ which acts as a regulator in the IR. The CP is characterized by $\xi\rightarrow\infty$. The fluctuation of $\xi$ are not measured directly. However, these fluctuations do influence fluctuations of observed particles, e.g., photons, to which the critical mode couples. We consider the glueball field $\chi (x)$ as the stochastic one obeying the Langevin-type equation
\begin{equation}
\label{e2}
 \frac{\partial L [\chi(x)]}{\partial\chi (x)}  =  h(x), 
\end{equation}
where $h(x)$ is the random (stochastic) source.
Because of the gluon fluctuations influence, $h(x)$ can be given as $h(x) = m_{g}B_{\mu} (x) I^{\mu}$, where $B_{\mu}$ stands as the vector (gluon) field, $m_{g}$ is the effective thermal gluon mass and $I_{\mu}$ is an auxiliary unit vector. 

The probability distribution of the field $\chi$ is given by $\mathcal{P}[\chi] \sim\exp (-G[h]/T)$, where 
$$ G[h] = \ln\int \mathcal {D} \chi\,e^{-\int d x\, L(x)}$$
defines the free energy $F_{h}$ averaged over $h(x)$  (see [8])
$$ F_{h} = \int \mathcal {D} h\,G[h]\,e^{-\frac{1}{2}\int d x\,h^{2}(x)} $$
with the classical LD
$$L (x) = \frac{1}{2} \chi(x)\,\Delta \chi (x) - \frac{1}{2} m^{2}\chi^{2} (x) - \lambda\,\chi ^{4}(x) + h(x)\chi (x).$$
The following expression averaged over $B_{\mu}$ 
$$ {\langle B^{a}_{\mu}(x)\,B^{b}_{\nu}(x^{\prime})\rangle} = 
 2\delta^{ab}\,\delta_{\mu\nu}\,\delta (x - x^{\prime})\,$$
is the Gaussian color noise because of the presence of color field. 
The most IR divergent contributions are from the maximum number of $h^{2}(x)$ insertions. 
For a sufficiently large ratio $m_{g}/T$  near the phase transition the gluon thermodynamic potential 
$$\Omega_{g}\sim T\int\frac{d^{3}\vec {p}}{(2\,\pi)^{3}} \ln \left (1-e^{-E_{g}/T}\right ),\,\,\, E_{g} = \sqrt {{\vert\vec {p}\vert}^{2} + m^{2}_{g}}$$
can be expanded with the strong effective gauge coupling $g(T) = m_{g}(T)/T $.

The two-point correlation  function (TPCF) is 
\begin{equation}
\label{e3}
W_{h}(x) ={\langle \chi (x)\chi(0)\rangle}_{h} \sim \xi^{4}\int\mathcal{D} h\,\tilde\chi_{h} (x)\,\tilde\chi_{h}(0)\,e^{-\frac{1}{2}\int d y h^{2} (y)}, 
\end{equation}
where $\chi (x) = \xi^{2}\tilde\chi_{h} (x)$ is the solution of Eq.({\ref{e2}) with the strong enough gluon stochastic source $h(x)$, where
$$\tilde\chi_{h} (x) \simeq \frac{1}{V} e^{( x + const)/\xi} - h(x), $$
$V$ is the volume conjugate to the scale of IR. The most significant result of (\ref{e3}) is the strong dependence on $\xi$ which is one of very sensitive signatures of the CP as $\xi\rightarrow \infty$. The maximum of the probability distribution $\mathcal{P} [\chi]$ is at $\chi (x) =0$, where the strong distortion due to gluon fluctuations at large distances $x\simeq \xi\,\ln(h V) - const$,  are occurred.
We find that the almost massless dilatons mediate the long-range strong forces which influence on the global dynamics above the phase transition. On the contrary, the deconfined phase is provided by the color charge screening at distances of the order of Debye mass.

3. A light dilaton is a natural object in the theory with an approximate scale symmetry accompanied by the spontaneous symmetry breaking of chiral symmetry. There is no such state in QCD. 
The production of $\chi$ would be via two gluon fusion, $gg\rightarrow\chi$. The corresponding cross section in proton-proton collisions is
$$\sigma (pp\rightarrow \chi\chi) = \sigma (pp\rightarrow h X)\frac{\Gamma (\chi\rightarrow gg)}{\Gamma (h\rightarrow gg)}, $$
where $\Gamma$ stands for the partial decay width and $h$ is the light hadron, e.g., $\pi^{0}$ meson. The dilatons are unstable and they decay into two photons. At high $T$ the CA acts as the source of primary photons (or soft photons in heavy-ion collisions [9]) not produced in the decays of hadrons.
When $T$ is going down the $\chi$ states disappear and there will be an abundant production of photons through the decays of light hadrons.

4. The couplings of $\chi$ to the Standard Model (SM) fields are easily be established using the Higgs-like models by replacing the vacuum expectation value of the Higgs by $f$ subject to $\chi$:
$$\partial_{\mu} \langle 0\vert S^{\mu} (x)\vert \chi (p)\rangle = \langle 0\vert \theta^{\mu}_{\mu} (x)\vert \chi (p)\rangle = -f\,m^{2}\, e^{-i\,p\,x}, $$
where for the on-shell case
$$\langle 0\vert \theta^{\mu\nu} (x)\vert \chi (p)\rangle = f (p^{\mu} p^{\nu} - g^{\mu\nu} p^{2} ) e^{-i\,p\,x}, \,\,\, p^{2} = m^{2}, $$
and $\langle 0\vert$ is the vacuum state corresponding to spontaneously broken dilatation (and chiral) symmetry.

In the exact scale symmetry $\chi$ couples to SM particles through the trace of $\theta_{\mu\nu}$
\begin{equation}
\label{e4}
 L = \frac{\chi}{f} \left ( \theta^{\mu}_{\mu\,{tree}} +  \theta^{\mu}_{\mu\,{anom}}\right ).
\end{equation}
The first term in (\ref{e4}) is proportional to particle masses
$$\theta^{\mu}_{\mu\,{tree}} = -\sum_{q} [m_{q} + \gamma_{m}(g) ]\bar q q + 2 m^{2}_{W} W^{+}_{\mu} W^{{-}\,\mu} + m^{2}_{Z} Z^{\mu} Z_{\mu} - \frac{1}{2} m^{2}\chi^{2} +\partial_{\mu}\chi\partial^{\mu}\chi, $$
 where $q$ are quark d.o.f. with the mass $m_{q}$, $\gamma_{m}$ are the corresponding anomalous dimensions. In contrast to the SM, the dilaton couples to massless gauge bosons even before running any SM particles in the loop, through the trace anomaly.
The latter has the following term for photons and gluons:
$$\theta^{\mu}_{\mu\,{anom}} = -\frac{\alpha}{8\,\pi}\, b_{EM}\, F_{\mu\nu}F^{\mu\nu} - 
\frac{\alpha_{s}}{8\,\pi}\sum_{i}\, b_{0_{i}}\, G_{\mu\nu}^{a}G^{{\mu\nu\,a}},$$
where $\alpha$ is the fine coupling constant, $b_{EM}$ and $b_{0_{i}}$ are the coefficients of electromagnetic (EM) and QCD $\beta$ functions, respectively. If the strong (and EM) interactions are embedded in the conformal sector the following relation for light and heavy particles sectors is established above the scale $\Lambda$ (in UV) [5]: 
$\sum_{light} b_{0} = - \sum_{heavy} b_{0}$, where the mass of $\chi$ splits the light and heavy states. The anomaly (non-perturbative) term for gluons in (\ref{e4})
$$\frac{\alpha_{s}}{8\,\pi}\,b^{light}_{0}  G_{\mu\nu}^{a}G^{{\mu\nu\,a}} = \frac{\beta (g)}{2\,g} G_{\mu\nu}^{a}G^{{\mu\nu\,a}}, \,\,\, b^{light}_{0} = -11 +\frac{2}{3}n_{L} $$
is evident, where the only $n_{L}$ particles lighter than $\chi$ are included in the $\beta$-function, $\beta (g) = b^{light}_{0}\,g^{3}/(16\, \pi^{2})$. For $m\sim O(\Lambda)$ one has $n_{L} = 3$ that indicates about 14 times increase of the $\chi\,g\,g$ coupling strength compared to that of the SM Higgs boson.

5. The light dilaton operates with low invariant masses where two photons are induced effectively by gluon operators. In the low-energy effective theory, valid below the conformal scale $\Lambda_{conf} = 4\,\pi\,f$, at small transfer-momentum $q$  we have
$\langle \gamma\gamma\vert \theta^{\mu}_{\mu} (q)\vert 0\rangle\simeq 0$  and [10] 
$$\langle \gamma\gamma\vert \frac{b^{light}_{0}\,\alpha_{s}}
{8\,\pi} G^{a}_{\mu\nu} G^{\mu\nu\,a}\vert 0\rangle = - \langle \gamma\gamma\vert \frac{b_{EM}\,\alpha}{8\,\pi} F_{\mu\nu} F^{\mu\nu}\vert 0\rangle, \,\, \vec q = 0.$$
The partial width  of the decay $\chi\rightarrow\gamma\gamma$ is
$$ \Gamma (\chi\rightarrow\gamma\gamma)\simeq \left (\frac{\alpha\,F_{anom}}{4\,\pi}\right )^{2}\,\frac{m^{3}}{16\,\pi\,f^{2}},$$
where the only CA does contribute through 
$$F_{anom} = -(2\,n_{L}/3) (b_{EM}/b^{light}_{0}), \,\,\,b_{EM} = -4\sum_{q:u,d,s} e^{2}_{q} = -8/3, $$ 
$e_{q}$ is the charge of the light quark. In the vicinity of IRFP the mass of $\chi$ is given by $m\simeq \sqrt{1-N_{f}/N^{c}_{f}}\Lambda$ [3], where $N^{c}_{f}$ is the critical value of $N_{f}$ corresponding to $\alpha^{c}_{s}$ at which the chiral symmetry is breaking and the confinement is appeared. In order to estimate $\Gamma(\chi\rightarrow\gamma\gamma)$ we take $f\simeq\Lambda$. 
When one approaches the CP the absolute value of $F_{anom}$ decreases due to increasing of $b^{light}_{0}$ as $n_{L}\rightarrow 0$. The second-order phase transition is characterized by the limits $N_{f}\rightarrow N^{c}_{f}$ and $\Lambda\rightarrow 0$, hence no primary photons should be evident through a detector. In the IR ($\alpha^{\star}_{s} > \alpha^{c}_{s}$) one can estimate the rate of primary photons escape compared to the width of the decay $\pi^{0}\rightarrow \gamma\gamma$ [11]: 
$$R_{\chi} = \frac{\Gamma(\chi\rightarrow\gamma\gamma)}
{\Gamma(\pi^{0}\rightarrow\gamma\gamma)} \sim 4 \%.  $$
Actually, $R_{\chi}\rightarrow 0$ as $\alpha^{\star}_{s}\rightarrow \alpha^{c}_{s}$ at the CP.
On the other hand, $R_{\chi} = 0$ should be viewed as an UV fixed point of the strong coupling phase transition.
 Thus one expects the suppression of primary photons escape once going away from IR to UV.

6. In summary, the novel approach to an approximate scale symmetry breaking up to the phase transition at the critical (end) point is suggested. The possible determination of the phase boundary between the confinement-deconfinement border and the high $T$ plasma phase can be seen inside the conformal window.
We find the singular behavior of TPCF (\ref{e3}) $\sim\xi^{4}$ as the CP is approached ($\xi\rightarrow\infty$).
The CP can be found as those followed by IRFP where the primary photons are detected. The origin of these photons is CA through the decays of the dilatons. When the incident energy scans from high to low values, the deviation of the primary photons escape rate $R_{\chi}$ from zero to a couple percents (primary photons production compared to that from  $\pi^{0}\rightarrow\gamma\gamma $ decay)  will indicate the appearance of CP. At the CP no escape of the primary photons are seen.

\end{document}